\documentclass[preprint]{aastex}

\usepackage{color,amsmath}
\newcommand\oops[1]{{#1}}
\newcommand\Varm{\mbox{$V_{\rm arm}$}}
\newcommand\Vwav{\mbox{$\langle V_w \rangle$}}
\newcommand\kmps{\mbox{$\rm km\,s^{-1}$}}
\newcommand\Msun{\mbox{$M_\sun$}}
\newcommand\Mspy{\mbox{$\Msun\rm\,yr^{-1}$}}

\shorttitle{Spiral model for CIT 6}
\shortauthors{KIM et al.}

\begin{document}
\title{Evidence of a Binary-Induced Spiral from an Incomplete Ring Pattern of CIT 6}
\author{Hyosun Kim\altaffilmark{1}}
\author{I-Ta Hsieh\altaffilmark{1}}
\author{Sheng-Yuan Liu\altaffilmark{1}}
\author{Ronald E. Taam\altaffilmark{1,2}}
\altaffiltext{1}{Academia Sinica Institute of Astronomy and Astrophysics, 
  P.O. Box 23-141, Taipei 10617, Taiwan; hkim@asiaa.sinica.edu.tw}
\altaffiltext{2}{Department of Physics and Astronomy, Northwestern University,
  2131 Tech Drive, Evanston, IL 60208}

\begin{abstract}
With the advent of high-resolution high-sensitivity observations, spiral 
patterns have been revealed around several asymptotic giant branch (AGB) 
stars. Such patterns can provide possible evidence for the existence of 
central binary stars embedded in outflowing circumstellar envelopes. Here, 
we suggest the viability of explaining the previously observed incomplete 
ring-like patterns with the spiral-shell structure due to the motion of 
(unknown) binary components viewed at an inclination with respect to the 
orbital plane. We describe a method of extracting such spiral-shells from 
an incomplete ring-like pattern to place constraints on the characteristics 
of the central binary stars. The use of gas kinematics is essential in 
facilitating a detailed modeling for the three-dimensional structure of the 
circumstellar pattern. We show that a hydrodynamic radiative transfer model 
can reproduce the structure of the HC$_3$N molecular line emission of the 
extreme carbon star, CIT 6. This method can be applied to other sources 
in the AGB phase and to the outer ring-like patterns of pre-planetary 
nebulae for probing the existence of embedded binary stars, which are 
highly anticipated with future observations using the Atacama Large 
Millimeter/submillimeter Array.
\end{abstract}


\section{INTRODUCTION}\label{sec:int}

Binary motion is now well known to generate a spiral pattern in the 
circumstellar envelope of \oops{a} mass losing star \citep{sok94,mas99,
he07,edg08,kim11,kim12a,kim12b,kim12c}. However, it is not generally 
recognized that the binary-induced pattern, vertically extended from 
the orbital plane, exhibits a ring-like pattern with an inclined 
viewing angle. At an inclination close to edge-on, in particular, 
the binary-induced spiral-shell appears as eccentric arcs in the 
plane of the sky \citep[see e.g.,][]{mas99,he07,kim12b}. Even at 
a low inclination (close to face-on), the pitch angle of the spiral 
pattern decreases with radius from the center of binary orbital motion 
\citep{kim12b}, such that the outer part of the spiral approximates a 
concentric ring structure. This misidentification may occur, especially, 
for the outer ring-like patterns of (pre-)planetary nebulae (PPN/PN) 
whose inner parts are contaminated by the bright, fast-moving lobes 
(e.g., NGC 7027 \citep{bon96}, Egg Nebula \citep{sah98}). Because of 
the incompleteness of the observed ring-like patterns, whether they 
arise from spherical shells due to the pulsation of single stars or 
spiral-shell patterns due to binary orbital motion is inconclusive. 
In this paper, we focus on the interpretation that the incomplete 
ring-like patterns observed in the past may be caused by the motion 
of (unknown) binary stars. 

Linking a binary scenario with the ring-like patterns observed on 
the circumstellar envelopes of asymptotic giant branch (AGB) stars and 
their remnant material in the outer parts of PPN/PN has an interesting 
implication for the late phases of stellar evolution. Since the bipolar 
(often multi-polar) structures, including jets and tori, observed in the 
PPN/PN \citep[e.g.,][]{bal02,sah11} likely originate from the central 
binary systems \citep{mor87,hug07,dem09}, a fundamental issue remains 
concerning the origin of the ring-like patterns in the outer parts of the 
PPN/PN. Such a ring-like pattern is commonly assumed as the consequence 
of the periodic mass loss of a {\it single} star in the previous AGB phase, 
due to stellar pulsation \citep[e.g.,][]{wil79,woo79} or thermal pulsation 
\citep[e.g.,][]{mat07} on the spherically symmetric wind. 
For example, the archetypal AGB star, IRC+10216, reveals the circumstellar 
pattern in the optical and infrared images \citep{lea06,dec11}, which is  
characterized by an interval corresponding to $\sim$\,250--1700 years. 
Interestingly, \citet{dec11} noted that this pattern interval of IRC+10216 
does not agree with either the stellar pulsation \citep[649 days,][]{leb92} 
or the thermal pulsation \citep[6000--33000 years,][]{lad10}. 
Similar discrepancy has also been found in several PPN/PN \citep{su04}.
A binary-induced spiral-shell model, in which the pattern interval time is 
defined by the binary orbital period without a strong limitation on the 
scale, may possibly serve as an alternative explanation for the ring-like 
(but not perfectly symmetric) patterns formed in the AGB phase. 

In the binary model, the observed spiral-shell pattern \oops{of a 
circumstellar envelope} can constrain the binary mass and orbital 
properties. For the carbon-rich AGB star, AFGL 3068, displaying a 
well-defined spiral circumstellar pattern in dust scattered light, 
\citet{mau06} estimated the orbital period simply from the arm 
interval and wind expansion speed, ignoring the inclination. For 
another carbon star, R Sculptoris, exhibiting a spiral pattern in 
addition to a thin thermal pulse shell, \citet{mae12} characterized 
the thermal pulsation with the help of their modeled binary properties 
for the spiral pattern features. 
Similarly, the circumstellar patterns in close binary systems may be used 
in determining the relative importance and observational consequence of the 
accretion disks surrounding the companion star, as theoretically investigated 
\citep[e.g.,][]{the93,nor06}. Furthermore, \citet{hua13} found that the bow 
shock around a disk can lead to complicated time-dependent post-shock 
structures for orbital separations less than about 20 AU. 
As such, the features of a spiral-shell pattern provide indirect evidence 
for the presence of a binary companion to the AGB star. In cases where the 
companion is obscured by the circumstellar envelope, the detection of spiral 
features may be the only diagnostic probe for, otherwise hidden binary systems.

In a previous paper \citep{kim12c}, we took account of the effect of 
inclination angle, for the first time, in modeling an observed circumstellar 
pattern within the binary framework. It was shown that the elongated spiral 
shape in the plane of the sky, seen in dust scattered light, can be used to 
constrain key binary quantities (i.e., inclination of the orbital plane, 
orbital period, companion mass, and mass ratio). However, this analysis left 
a degeneracy in such model parameters since the apparent shape in the plane 
of the sky reflects both the binary motion and projection effect due to the 
inclination of the orbital plane. In this paper, we show that the degeneracy 
can be lifted by using gas kinematics from high resolution observations, which 
provide three-dimensional information. To illustrate the modeling method, we 
apply the model to the  object, CIT 6, by revisiting its incomplete ring-like 
pattern observed in molecular line emission with the spiral-shell due to the 
motion of (hypothesized) central binary stars. 

In Section \ref{sec:cit}, we introduce the known properties of CIT 6 and 
the existing molecular line observation with the Very Large Array (VLA), 
which are used to constrain the model parameters. 
In Section \ref{sec:geo}, we formulate a simple analytic model based on 
the oblate spheroid geometry of an Archimedes spiral-shell with a binary 
orbital inclination $i$. This simple analytic model can be easily applied 
to other sources in order to ascertain whether a spiral-shell pattern exists. 
In Section \ref{sec:deg}, a parameter space analysis is carried out following 
the method of \citet{kim12c} based on the elongated shape of the pattern in 
the plane of the sky. 
In Section \ref{sec:hyd}, \oops{the results of} hydrodynamic radiative 
transfer simulations are \oops{presented, from which we use} molecular line 
kinematics to resolve the degeneracy in the parameter space analysis. We 
find that the simulated spiral-shell model can reproduce the observational 
data in the channel maps and position-velocity (P-V) diagrams, providing 
a better fit than obtained by using a spherically symmetric shell model. 
In Section \ref{sec:dis}, we discuss the advantage and limitation of the 
current model. Finally in Section \ref{sec:sum}, we summarize our findings 
and suggest future applications.

\section{KNOWN \oops{FACTS} OF CIT 6 AS MODEL CONSTRAINTS}\label{sec:cit}

CIT 6 (also known as RW LMi, IRC+30219, IRAS 10131+3049, AFGL 1403) is an 
extreme carbon star and believed to be in the transition from the AGB to 
the post-AGB phase. \citet{sch02} found the presence of a nascent bipolar 
nebula, providing evidence that the evolutionary phase of CIT 6 lies just 
pass the tip of AGB. Its effective temperature of 2800\,K \citep{coh79} 
and the bolometric luminosity of $\sim10^4\,L_\sun$ \citep{lou93} also 
correspond to the values at the tip of AGB phase of the evolutionary track 
for a star with the initial mass of 2--3\,\Msun, according to the Single 
Stellar Evolution package of \citet{hur00} assuming a solar metallicity. 
The distance to CIT 6 is estimated to be $400\pm50$\,pc based on the 
pulsation period ($1.65\pm0.02$ yr) using a period-luminosity relation 
\citep[and references therein]{coh96}. 

The importance of CIT 6 stems from the existence of partial rings of molecular 
line emission based on VLA interferometric observational data obtained by 
\citet[\oops{see also the red-scaled image in Figure\,\ref{fig:chm}}]{cla11} 
at high angular resolution ($\sim0.\!\!\arcsec7$) and high spectral 
resolution ($\sim1\,\kmps$). \citet{cla11} 
modeled this line emission pattern with four spherical shells, using 
the shell radii, expansion velocities, and central velocities as free 
parameters for the individual shells. Their model fitting results in 
different central velocities between the modeled shells, which may imply 
the departure from spherical symmetry of the circumstellar pattern
about the central AGB star. \citet{cla11} remarked on the existence 
of the asymmetric features such as the spiral shape (relatively well 
traced in the velocity channel of $-7.4\,\kmps$), stronger emission 
observed in the redshifted side, and lack of emission in the west and 
northwest near the central channels. \citet{cha12} showed that the 
optical depth effect cannot reproduce the asymmetric line profile of 
CIT 6 for a physical structure assumed to be spherically symmetric. 
Their three-dimensional morphokinematic model instead suggested 
several incomplete shells for the structure of CIT 6. 

\oops{Here w}e suggest an alternative interpretation for the asymmetric 
shape of the circumstellar pattern as an indication for a central binary 
star system. Some evidence for binarity is indeed found in the Hubble 
Space Telescope (HST) images with two compact cores separated by 
$0.\!\!\arcsec17$ at the position angle (PA, measured from north to 
east) of 10\arcdeg\ \citep{mon00}. The northern lobe varies in its 
flux and polarization with the known pulsation period of the carbon 
star ($1.65\pm0.02$ yr; \citealp{coh96}), while the bluer southern lobe 
is stable and has a polarization direction $\sim90\arcdeg$ different from 
the northern red component \citep{tra97}. These characteristics of the 
southern lobe, differing from those of the carbon star and the northern 
lobe, suggest that the blue southern lobe is illuminated by a companion 
star in the vicinity \citep{mon00}. The existence of \oops{a} companion 
star is supported by the consistence of the polarization direction of 
H$_\alpha$ emission \citep{tra94} to the blue component. From the observed 
flux at 4500\AA, \citet{sch02} inferred that the companion is likely later 
than spectral type G0 ($\ga1$\,\Msun) if it is a main-sequence star.

\section{ANALYTIC MODEL USING GEOMETRY OF INCLINED ARCHIMEDES SPIRAL-SHELL}
\label{sec:geo}

The orbital motion of a mass losing star in a binary system creates a 
spiral-shell pattern in its circumstellar envelope \cite[e.g.,][]{sok94}. 
We consider the overall morphology of the spiral-shell pattern as an 
oblate spheroid, following \citet{kim12c}. In the coordinates having 
the vertical axis $z^\prime$ aligned with the orbital axis, the oblate 
spheroid can be described by (see Figure\,\ref{fig:geo})
\begin{equation}\label{eqn:obl}
  (x^\prime/a)^2+(y^\prime/a)^2+(z^\prime/b)^2=1\oops{,}
\end{equation}
where the axial ratio $a/b$ is defined by the ratio of pattern propagation 
speeds $\Varm/\Vwav$ in the orbital plane and along the polar axis. Here, 
the radial propagation speed of the spiral pattern in the orbital plane is 
affected by the orbital motion of the mass losing star with the orbital 
speed of $V_1$ to be $\Varm=\Vwav+2V_1/3$ \citep{kim12b}, while the pattern 
propagation speed along the orbital axis is the average wind speed $\Vwav$ 
corresponding to the intrinsic wind speed.

We define new coordinates $(x,\,y,\,z)$ rotated by an inclination $i$, 
in which the observer is at infinity in the direction of the $z$-axis. 
A rotational transformation about the $x$-axis by the angle $i$ is defined by 
$x^\prime=x$, $y^\prime=y\cos i+z\sin i$, and $z^\prime=-y\sin i+z\cos i$. 
Substitution of these into Equation (\ref{eqn:obl}) yields
\begin{equation}\label{eqn:inc}
  R^2(1+f\sin^2i\,\sin^2\phi) + z^2(1+f\cos^2i) = a^2,
\end{equation}
where $R$ and $\phi$ refer to the tangential coordinate $R=(x^2+y^2)^{1/2}$ 
and the azimuthal angle in the $x$--$y$ plane, respectively, defining 
$x=R\cos\phi$  and $y=R\sin\phi$. A factor $f$ is defined as $f=(a/b)^2-1$, 
thus 
\begin{equation}
  f = (\Varm/\Vwav)^2-1.
\end{equation}
Alternative forms of Equation (\ref{eqn:inc}) are
\begin{equation}
  R=a (1+f\sin^2i\,\sin^2\phi)^{-1/2} \left(1-(z/a)^2(1+f\cos^2i)\right)^{1/2},
\end{equation}
or
\begin{equation}
  z=a (1+f\cos^2i)^{-1/2} \left(1-(R/a)^2(1+f\sin^2i\,\sin^2\phi)\right)^{1/2}.
\end{equation}
Because the oblate spheroid geometry originates from the latitudinal 
variation of the wind velocity, these coordinates $R$ and $z$ can be 
linearly transformed to the tangential and line-of-sight velocities of 
the wind. Thus, the tangential component is 
\begin{eqnarray}\label{eqn:v_t}
  V_t = && \Varm \times ( 1+f\sin^2i\,\sin^2\phi )^{-1/2} \nonumber\\
  && \times ( 1-(V_r/\Varm)^2(1+f\cos^2i) )^{1/2},
\end{eqnarray}
representing the pattern propagation speed in the inclined plane within 
the $V_r$-channel. The second term in the right-hand-side of Equation 
(\ref{eqn:v_t}) shows the inclination effect, and the last term is related 
to the channel velocity representing the height of the relevant material 
from the midplane. From Equation (\ref{eqn:v_t}), the line-of-sight velocity 
can be written as 
\begin{eqnarray}
  V_r = && \Varm \times (1+f\cos^2i)^{-1/2} \nonumber\\
  && \times ( 1-(V_t/\Varm)^2(1+f\sin^2i\,\sin^2\phi) )^{1/2},
\end{eqnarray}
at a given $V_t$. The maximum line-of-sight velocity is 
\begin{equation}\label{eqn:vrm}
  (V_r)_{\rm max} = \Varm \times ( 1+f\cos^2i )^{-1/2},
\end{equation}
at the orbital center (i.e., at the center of mass in the circular orbit case).

A circumstellar spiral pattern due to the orbital motion of the AGB star 
\oops{(orbital radius $r_1$; orbital velocity $V_1$) for a given $z$} is 
defined by a differential equation in the polar coordinates $(R,\,\phi)$ as,
\begin{equation}\label{eqn:dif}
  \frac{1}{r_1}\frac{{\it d}R}{{\it d}\phi}=\frac{V_t}{V_1},
\end{equation}
where the tangential velocity $V_t$ represents the pattern propagation 
velocity in the layer \oops{within the $V_r$-channel. The velocity 
$V_t$} is given by Equation (\ref{eqn:v_t}) as a function of $\phi$ 
with given wind velocity, orbital velocity, and inclination angle. 
As the velocity of an AGB wind is nearly constant beyond the wind 
acceleration region (typically less than several times the stellar 
radius; \citealp{hof07}), the integrated form of \oops{E}quation 
(\ref{eqn:dif}) well approximates to 
\begin{equation}\label{eqn:pol}
  \left(\frac{R}{1\arcsec}\right)
  = \left(\frac{V_t}{V_1} \right) \left(\frac{r_1}{1\rm AU}\right)
  \left(\frac{d}{1\rm pc}\right)^{-1} \left(\frac{\phi}{1\rm rad}\right)\oops{,}
\end{equation}
for the pattern located at a large radius $R$.
Here, $d$ is the distance to the source. 

\section{PARAMETER SPACE ANALYSIS}\label{sec:deg}

In order to obtain approximate constraints on the binary 
properties, we explore the possible parameter space for the system 
of interest, following the method described in \citet{kim12c} using 
an overall oblate spheroidal morphology of the pattern. Four observed 
quantities are required, corresponding to the projected axial ratio 
$(a/b)_{\rm proj}$, the arm pattern spacing $\Delta r_{\rm arm}$ along 
the major axis, the projected binary separation $(r_1+r_2)_{\rm proj}$, 
and the angular position of the stars relative to the major axis of the 
elliptical pattern shape $\phi_{\rm binary, proj}=\rm PA_{binary, proj}
-PA_{node}$. Here, $\rm PA$ refers to a position angle measured from 
the north to the east direction and hereafter $\phi$ is defined as the 
angle relative to the line of nodes of the binary orbit ($\rm PA_{node}$; 
the intersection of the orbital plane with the plane of the sky) measured 
in the counterclockwise direction. Refer to \citet{kim12c} for the details 
of this method using an overall oblate spheroid shape. 

We first identify the line of nodes of the binary orbit as the major 
axis of the elongated spiral pattern of the HC$_3$N(4--3) line emission 
in Figure\,\ref{fig:chm}. Because of the insufficient sensitivity 
of observation, it is challenging to precisely determine the angular 
direction of the major axis only by ellipse fitting. Thus, we also use 
the fact that the P-V diagram is most symmetric along the line of nodes 
and most asymmetric in the perpendicular direction (middle and right 
panels of Figure\,\ref{fig:pvd}, respectively; see also Section 
\ref{sec:inc}). From the resulting ellipse fit with the major axis at 
$\rm PA_{node}=10\arcdeg$, we measure the axis ratio $(a/b)_{\rm proj}$ 
of 1.15 and the pattern spacing $\Delta r_{\rm arm}$ of $3.\!\!\arcsec2$ 
along the major axis. We assume the two compact components from the 
optical images are members of a binary (or their surrounding gaseous 
envelopes) following \citet{mon00}, which provides a projected separation 
of $(r_1+r_2)_{\rm proj}=0.\!\!\arcsec17$ and a position angle of the 
binary at $\rm PA_{binary, proj}=10\arcdeg$. The line of nodes 
nearly coincides with the alignment of the two stars, implying that 
$(r_1+r_2)_{\rm proj}$ is approximately equivalent to the actual binary 
separation. In addition to these four observed quantities, the distance
of 400\,pc is adopted (see Section \ref{sec:cit}), thus constraining 
the binary properties in a parameter space spanning the average wind 
velocity $\Vwav$ versus the orbital velocity $V_1$ of the mass losing 
star (Figure\,\ref{fig:deg}).

The gray-colored area in Figure\,\ref{fig:deg} shows the possible sets of 
binary parameters. The first constraint for the gray area is the mass of 
the evolved carbon star ($M_1$, black solid line) greater than the mean 
mass of white dwarfs \citep{mad04}. In particular, the stellar luminosity 
of $\sim10^4\,L_\sun$ for CIT 6 suggests the core mass of $\sim0.69\Msun$ 
with 2\% accuracy based on \citet{pac71}, providing the lower limit 
on the current mass, $M_1$, of the AGB star. The other constraint stems 
from the maximum line-of-sight velocity of the pattern $(V_r)_{\rm max}$ 
corresponding to the half width of the HC$_3$N(4--3) line, measured as 
18\,\kmps\ with an uncertainty of $\pm0.5\,\kmps$ considering the spectral 
resolution. In previous papers \citep{kim12b,kim12c}, the lower limit of 
the maximum line-of-sight velocity was loosely constrained by the condition 
$(V_r)_{\rm max}\la\Vwav+V_1$, which is satisfied at any inclination. 
We further calculate the maximum line-of-sight velocity taking account 
of the inclination effect. From equation (\ref{eqn:vrm}), the measured 
line-of-sight velocity range $(V_r)_{\rm max}=18\pm0.5\,\kmps$ is marked 
in Figure\,\ref{fig:deg} as \oops{a set of} yellow dotted lines.

The plausibility of a particular model parameter set can be easily checked 
by the simple analytic model in Section \ref{sec:geo}. Among the possible 
models in Figure\,\ref{fig:deg}, the parameter set marked by a plus symbol 
is applied to the analytic model. The model consists of an estimate of the 
AGB star mass of $M_1=0.8\,\Msun$, companion mass $M_2=2.2\,\Msun$, binary 
separation $r_1+r_2=68$\,AU, inclination $i=60\arcdeg$, and average wind 
velocity $\Vwav=15.7\,\kmps$ with the distance $d=400$\,pc. The result 
is overlaid in each velocity channel map of Figure\,\ref{fig:chm} by a 
gray-colored solid line, showing a reasonable match with the outer edge 
of the observed pattern (red-scaled), particularly for the middle-row 
panels displaying the near-central velocity channels. Furthermore, this 
inclined spiral can explain the wiggle feature of the observed pattern 
in the polar coordinates, clearly seen in Figure\,\ref{fig:pol}. Without 
considering the inclination effect (i.e., assuming face-on; $i=0$), the 
pattern propagation speed $V_t$ in each $V_r$-velocity channel is constant 
in Equation (\ref{eqn:v_t}), indicating that the pattern should appear as a 
straight line in the polar coordinates with the slope determined by Equation 
(\ref{eqn:pol}). For comparison, a concentric spherical shell pattern forms 
horizontal lines in the $\phi$-$R$ coordinates in Figure\,\ref{fig:pol}. 
Therefore, the wiggle feature, which is also found in other sources (e.g., 
R Sculptoris, see Figure 2a in \citealp{mae12}), is a natural outcome of 
the inclined spiral model, and is distinguishable from the face-on view 
of the spiral model for a binary system and the spherical shell model for 
a single star. 

\section{HYDRODYNAMIC RADIATIVE TRANSFER MODEL}\label{sec:hyd}

The region delineated by the gray-colored area in the parameter space 
(Figure\,\ref{fig:deg}), based on the elongated spiral shape projected 
on the plane of the sky, can be sharpened by taking advantage of three%
-dimensional information provided by the molecular line kinematics. We 
explore the parameter space in the vicinity of the $(V_r)_{\rm max}=18$%
\,\kmps\ line with hydrodynamic simulations followed by radiative transfer 
calculations for the HC$_3$N molecular line. The parameters for the model 
which best matches the aspects of the VLA observation \citep{cla11}, 
described in Sections \ref{sec:inc} and \ref{sec:res}, are summarized 
in Table\,\ref{tab:par} (and by a plus sign in Figure\,\ref{fig:deg}). 
We note that these parameters are not independent of each other due to 
their constraints imposed by the parameter space analysis.

\subsection{Numerical method}\label{sec:met}

The basic equations of hydrodynamics are integrated using the FLASH3 
code \citep{fry00} based on a piecewise parabolic method \citep{col84}. 
We construct a centrally concentrated, static grid utilizing the 
block-structured adaptive mesh refinement PARAMESH package and adjusting 
the refinement level based on the distance from the stars. The maximum 
refinement level of eight with the grids of $64\times64\times32$ per block is 
used for the simulation domain of 8400\,AU$\times$8400\,AU$\times$4200\,AU 
($21\arcsec\times21\arcsec\times10.\!\!\arcsec5$ at 400\,pc, corresponding 
to the image size of the VLA observation in consideration) in 
three-dimensional Cartesian coordinates, assuming the mirror symmetry about 
the orbital $x$--$y$ plane\footnote{Strictly speaking, the simulation domain 
lies in $(x^\prime,\,y^\prime,\,z^\prime)$ coordinates defined in Section 
\ref{sec:geo} and Figure\,\ref{fig:geo}. The prime notation is omitted 
from now on.}. The spatial gridding corresponds to a resolution of 1\,AU 
($0.\!\!\arcsec0025$) in the central region, assigning a sufficient number 
of grid cells at the wind generating radius $r_{\rm in}$ within which 
physical conditions are reset every simulational timestep, in order to 
assure the ejection of gas from the mass losing star in a spherical shape. 
The gas temperature over the simulation domain is scaled by the temperature 
$T_{\rm in}$ at the wind generating radius $r_{\rm in}$ from the mass losing 
star. We set $r_{\rm in}=\rm 10\,AU$, and $T_{\rm in}$ is calculated by the 
temperature distribution $T(r)=T_{\rm eff}(r/r_{\rm eff})^{-2(\gamma-1)}$ 
of an adiabatic gaseous medium with a $r^{-2}$ density distribution. 
The observed effective temperature $T_{\rm eff}=\rm 2800\,K$ is set 
at the assumed stellar photospheric radius $r_{\rm eff}=556\,R_\sun$ 
or 2.59\,AU. An adiabatic equation of state with the adiabatic index 
$\gamma$ of 1.4 is used.

The gravitational potentials of the stars are treated as Plummer potentials 
\citep[see, e.g.,][]{bin08} using the POINTMASS implementation of the FLASH3 
code with modifications to include the gravitational softening radii and the 
motion of the two stars. The gravitational softening radius of the mass losing 
star does not change the result as it is smaller than the wind generating 
radius. The gravitational softening radius of the companion is set to be 
comparable to the Hoyle-Lyttleton accretion radius. An additional simulation 
with the companion's softening radius reduced by a factor of 2 is compared 
with the model presented in this paper, in order to verify its insignificant 
contribution to the overall shape of the arm pattern. The accretion disk of 
the companion is expected to have a negligible size if it is gauged by the 
impact parameter (0.004 AU with the large separation of our binary system) 
in \citet{hua13}. Resolving the accretion disk for a wide binary system is 
computationally very expensive and beyond the scope of this paper. 

The FLASH3 results are used as inputs to the radiative transfer calculation 
for HC$_3$N lines with the SPARX code. SPARX, standing for Simulation Package 
for Astrophysical Radiative Transfer, is a software package for calculating 
molecular excitation and radiative transfer of (dust) continuum and molecular 
lines at millimeter/submillimeter wavelengths. It adopts the accelerated Monte 
Carlo (AMC) algorithm developed by \citet{hog00}, and finds full non-local 
thermodynamical equilibrium (non-LTE) solutions of molecular level population 
and radiation fields iteratively and consistently. Instead of assuming that 
the molecular excitation is simply in equilibrium corresponding to the local 
temperature, non-LTE demands only statistical equilibrium of the level 
population through collisional and radiative excitation and de-excitation 
processes, which can be solved with the detailed balance equation. Additional 
(de-)excitation mechanisms, such as infrared radiative pumping, have not been 
incorporated in the package. 

As a post-processing tool, the SPARX requires inputs of molecular gas 
density, temperature, velocity as well as the linewidth distribution, 
which can be provided by the results of a hydrodynamic simulation. 
Knowledge of the molecular abundance distribution and parameters such 
as dipole moments, line frequencies, and collisional rates are also 
required by the SPARX. The final SPARX output includes synthetic images 
which can be used for generating spectra and P-V diagrams, as input for 
a telescope filter for simulated observations, or for other scientific 
analysis. Details of the package such as the code implementation and 
its verification with benchmark cases will be presented in Liu et al. 
(2013, in preparation).

For the SPARX calculation, the HC$_3$N molecular parameters are obtained 
from the Leiden Atomic and Molecular Database (LAMDA) \citep{sch05}. 
The HC$_3$N abundance of $1.5\times10^{-6}$ is adopted throughout the 
envelope \citep{zha09}, while a spherical central region of 2\arcsec\ 
was carved out with HC$_3$N abundance set to zero in order to mimic the 
central hole seen in the observed emission of HC$_3$N (also found in 
its parent molecule CN; \citealp{lin00}). For computational efficiency, 
we use the spatial and spectral resolutions of $0.\!\!\arcsec13$ and 
0.4\,\kmps\ in the SPARX calculation, which are sufficient to simulate 
the VLA observation of \citet{cla11} summarized in Section\,\ref{sec:cit}. 
To further compare the SPARX output with the VLA observation, we perform 
a convolution with a Gaussian beam of $0.\!\!\arcsec7$ and the rebinning 
to 1\,\kmps\ over the synthetic images.

\subsection{Inclination effects on line kinematics}\label{sec:inc}

The inferred binary parameters are sensitive to the orbital inclination 
as illustrated in Figure\,\ref{fig:deg}. \oops{Hence, determination of 
the inclination can lift} the degeneracy that remain in the parameter 
space analysis based only on the pattern shape in the plane of the sky. 
With the binary parameters in Table\,\ref{tab:par}, Figure\,\ref{fig:inc} 
compares the images for the non-zero velocity channels and the P-V diagrams 
along the major (i.e., $x$-axis; line of nodes) and minor (i.e., $y$-axis; 
perpendicular to the line of nodes) axes for five different inclinations 
of 0\arcdeg\ (face-on), 30\arcdeg, 50\arcdeg, 70\arcdeg, and 90\arcdeg\ 
(edge-on). The channel images (top row) illustrate the change of pattern 
shape from a spiral in the face-on view to ellipses in the edge-on view. 
In particular, the spiral pattern is split along $-y$-axis in the image 
for $i=50\arcdeg$ and attached to the adjacent component in the image for 
$i=70\arcdeg$, making a new ring. 

The P-V diagrams of Figure\,\ref{fig:inc} (middle and bottom rows) reveal 
notable features. The middle row show that, regardless of inclination, the 
P-V diagram along the line of nodes (i.e., $x$-axis) appears relatively 
symmetric about the central velocity because of the mirror symmetry of the 
overall oblate spheroid geometry about the orbital plane. In contrast, the 
P-V diagram along the minor axis highly depends on the inclination angle 
(bottom row). One important implication is that the PA dependence of the 
P-V diagram can be used to determine the line of nodes of the binary orbit 
(as in Section \ref{sec:deg}). Moreover, the comparison between the P-V 
diagrams along major and minor axes helps to infer the inclination angle 
of the system from the observed data cube. For example, the PA dependence 
of the P-V diagram is minimized in the face-on case. 

The P-V diagram along the minor axis (bottom row) provides further 
information on the orbital inclination. At a mid-range inclination (see 
e.g. $i=30\arcdeg$), it reveals an asymmetry in intensity between the 
redshifted and blueshifted components, \oops{the} degree of asymmetry 
\oops{of which} differs with position, and thus the overall shape of P-V 
diagram is tilted. Hence, the degree of tilt and spectral asymmetry is 
dependent on the inclination. On the other hand, the maximum line-of-sight 
velocity (i.e., $V_r$ at $x=y=0$) increases with inclination as predicted 
in Equation (\ref{eqn:vrm}). This increase of maximum velocity makes the 
outline of P-V diagram at the high end velocity at each $y$ position 
($|V_r|\ga10\,\kmps$) shallower in the face-on case ($i=0\arcdeg$) and 
becoming steeper as the inclination increases. All these features of the 
P-V diagrams can be used to constrain the inclination angle of the binary 
system.

In the framework of \oops{the} binary model, the companion creates a 
gravitational\oops{ly-induced density} wake in a flatter spiral shape, 
overlapped with the spiral-shell pattern due to the orbital motion of 
the AGB star \citep{kim12b}. The overlapped region appears as knots in 
the plane of the sky at the PA depending on the inclination \citep[See 
Figure 10 of][]{kim12b} because the knots represent the companion's 
wake relatively thin in the vertical direction from the orbital plane 
\citep{kim12a}. In Figure\,\ref{fig:inc}, the top panels for a non-zero 
velocity channel show such knots at high inclinations ($i\ga50\arcdeg$ 
in this model) at the PA varying with the inclination (marked by arrows). 
In the P-V diagram along the minor axis (bottom row), knots appearing with 
low inclinations ($i\la50\arcdeg$) also shift the positions depending on 
the inclination. On the other hand, in the P-V diagram along the line of 
nodes of orbits (middle row) all knotty structures are present at the 
zero velocity, confirming the existence of the companion's wake on the 
orbital plane. 

In short, the shape of P-V diagram, the degree of spectral asymmetry, 
the position of knots both in individual channels and in individual P-V 
diagrams provide important clues in identifying the best model. It is also 
worth noting that the velocity dispersion of individual arm component is 
dominated by the velocity variation due to the binary orbital motion 
($\sim4.6\,\kmps$). For the binary system in Table\,\ref{tab:par} with 
an inclination angle of 60\arcdeg, the above features reasonably match 
the VLA observation.

\subsection{Fiducial model}\label{sec:res}

\begin{figure*} 
  \plotone{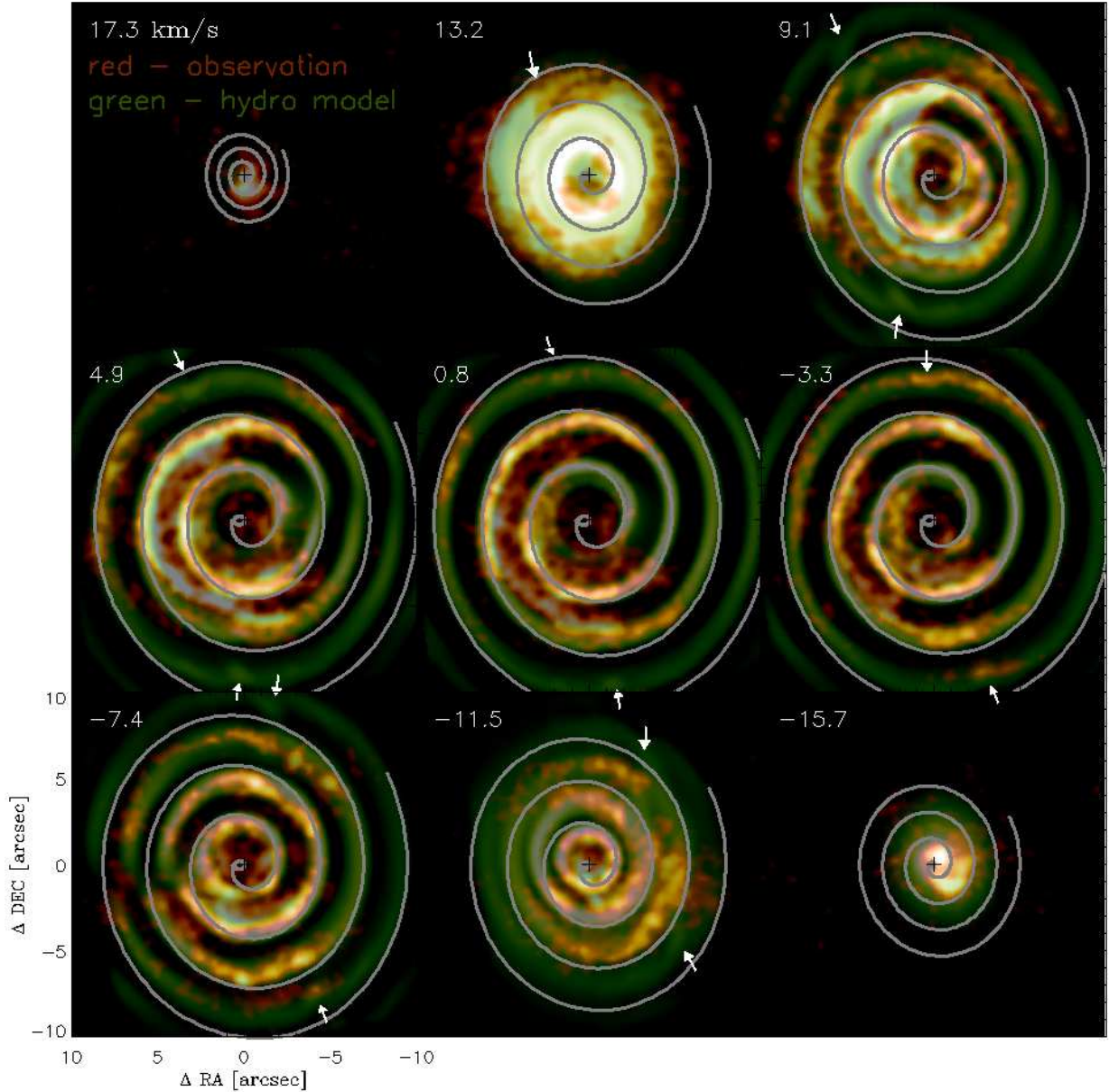}
  \caption{\label{fig:chm}
    Velocity channel images of the HC$_3$N(4--3) line emission of CIT 6 
    observed with the VLA ({\it red}, Figure\,1 of \citealp{cla11}), and 
    modeled by an hydrodynamic radiative transfer simulation using the 
    FLASH and SPARX codes ({\it green}). The red color ranges from 1 to 
    5\,mJy/beam logarithmically. In order to show the pattern shape in 
    the model beyond the detection limit of this particular observation, 
    the scale of the green color is displayed from 0.1 to 2\,mJy/beam 
    logarithmically with a synthetic beam size of $\sim0.\!\!\arcsec7$ 
    and velocity resolution of $\sim1\,\kmps$. North is up, and east is 
    left. The position angles of the knots found in the hydrodynamic model 
    are indicated by arrows. A simple geometric model with an inclined 
    Archimedes spiral-shell is drawn by a gray solid line in each channel. 
    The model parameters are summarized in Table\,\ref{tab:par}.
  }
\end{figure*}

\begin{figure} 
  \plotone{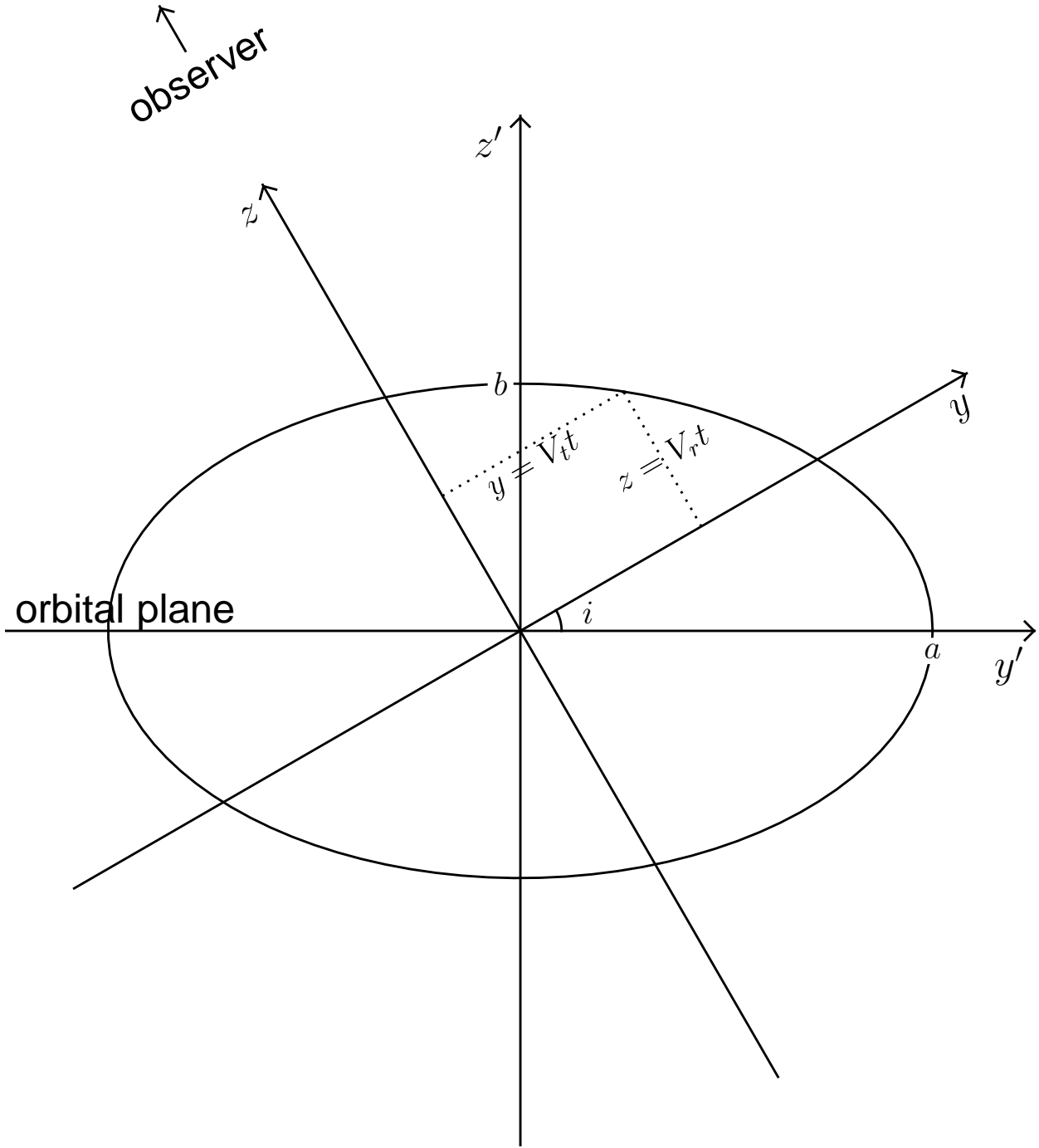}
  \caption{\label{fig:geo}
    A schematic diagram of an oblate spheroid viewed by an observer 
    with an inclination angle $i$, depicting the geometry in the prime 
    (intrinsic) and unprime (observational) coordinate frames.
  }
\end{figure}

\begin{figure*} 
  \plotone{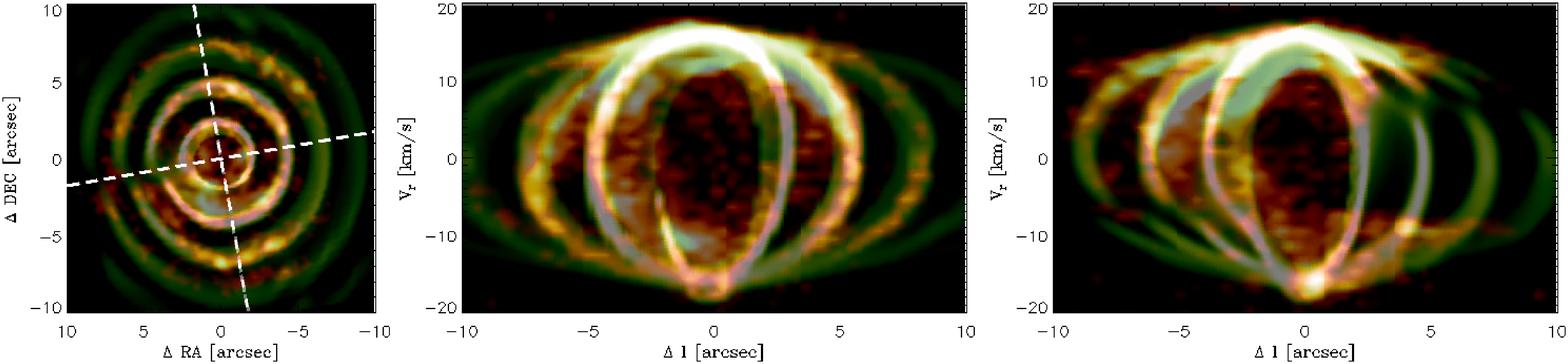}
  \caption{\label{fig:pvd}
    The HC$_3$N(4--3) line emission of CIT 6 taken with VLA ({\em red}, 
    \citealp{cla11}) compared with a spiral-shell model ({\em green}). 
    A channel map at $-7.4\,\kmps$ includes white dashed lines for the 
    major and minor axes at $\rm PA=10\arcdeg$ and 100\arcdeg\ ({\em left}), 
    along which the P-V diagrams present a relatively symmetric pattern 
    ({\em middle}, $\rm PA=10\arcdeg$) and the most asymmetric pattern 
    ({\em right}, $\rm PA=100\arcdeg$), respectively. The radius along 
    the major and minor axes, $\Delta l$, is measured from bottom to top 
    and left to right, respectively. The model parameters are tabulated 
    in Table\,\ref{tab:par}. The modeled orbit inclination is 60\arcdeg. 
    The red and green colors ranges from 1 to 5\,mJy/beam and 0.1 to 
    2\,mJy/beam, respectively, in a logarithmic scale. 
  }
\end{figure*}

\begin{figure*} 
  \plotone{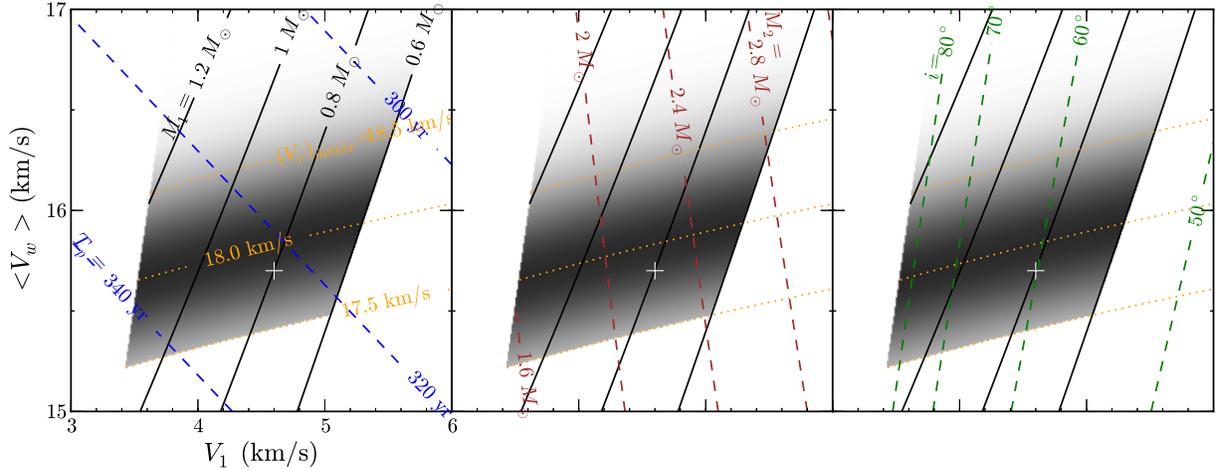}
  \caption{\label{fig:deg}
    Model parameter space of the undisturbed wind velocity $\Vwav$ versus 
    the orbital velocity $V_1$ of the mass losing star in a binary system, 
    based on the shape of an elongated spiral pattern with the projected 
    axial ratio of 1.15, the arm spacing of 1280 AU ($3.\!\!\arcsec2$ at 
    400\,pc), the binary stars aligned with the line of nodes of orbits 
    with the separation of 68 AU ($0.\!\!\arcsec17$). Black solid lines 
    represent the current mass $M_1$ of the mass losing star. Yellow dotted 
    lines show the half of linewidth of $18\pm0.5\kmps$, constraining the 
    model binary system within the gray-colored region favoring the darker 
    area. The binary orbital period, the companion mass $M_2$, the orbit 
    inclination $i$ are displayed by blue, red, and green dashed lines 
    in left, middle, and right panels, respectively. Plus sign corresponds 
    to a binary model for $M_1=0.8\,\Msun$, $M_2=2.2\Msun$, and $i=60\arcdeg$ 
    with the separation of 68\,AU, and the line-of-sight expansion velocity 
    of 17.8\,\kmps.
  }
\end{figure*}

\begin{figure*} 
  \plotone{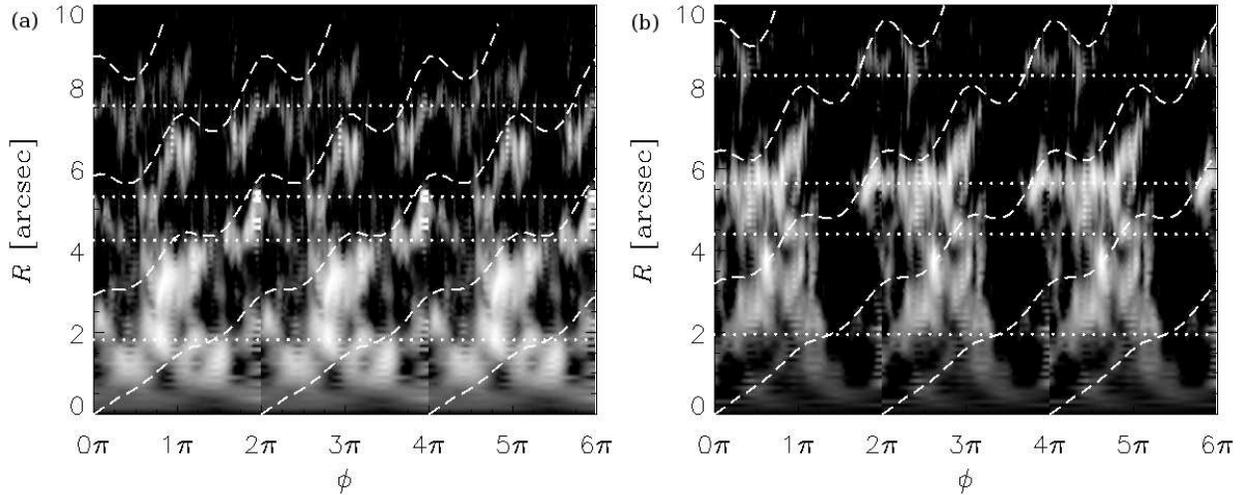}
  \caption{\label{fig:pol}
    The HC$_3$N(4--3) line emission in the channel velocity of (a) 
    $-7.4\,\kmps$ (left bottom in Figure\,\ref{fig:chm}) and (b) 
    $0.8\,\kmps$ (middle in Figure\,\ref{fig:chm}) plotted in the 
    polar coordinates. An elongated spiral model is drawn by dashed 
    lines, tracing the wiggle feature. For comparison, an Archimedes 
    spiral without the inclination effect (i.e., face-on) appears to 
    have a constant slope in the polar coordinates, connecting the 
    local peaks of the dashed lines. Dotted lines refer to the four 
    spherical shell model in \citet{cla11}. The gray color ranges 
    from 1 to 5\,mJy/beam in a logarithmic scale. 
  }
\end{figure*}

\begin{figure*} 
  \plotone{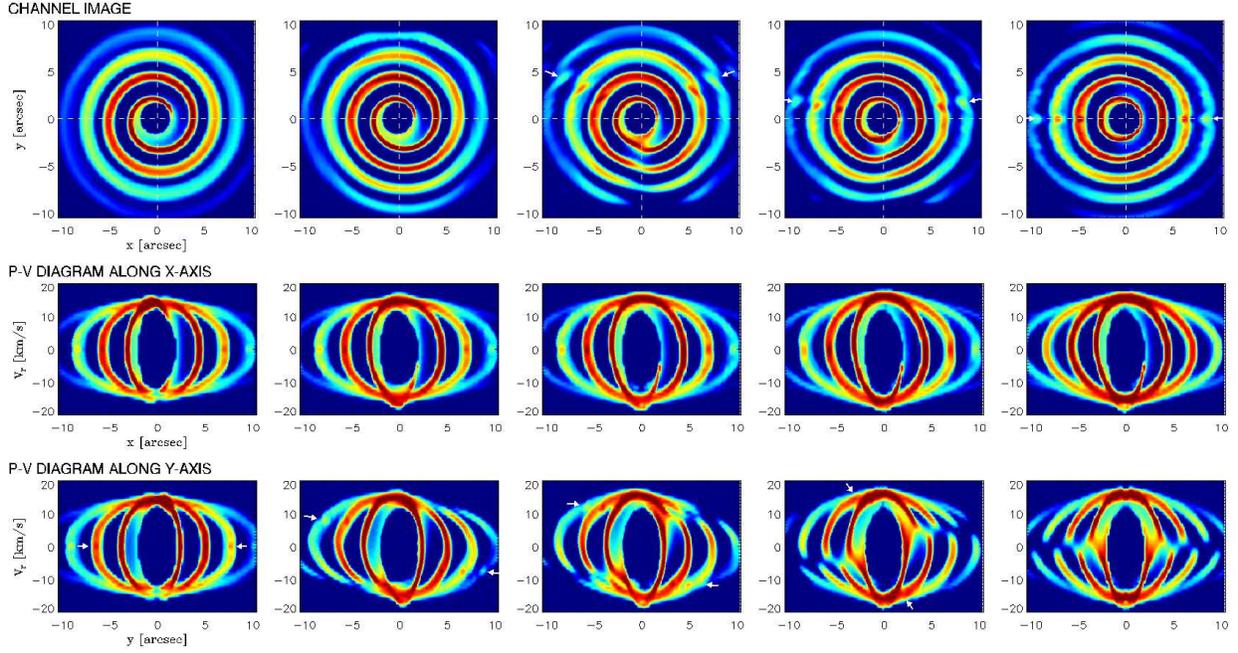}
  \caption{\label{fig:inc}
    The HC$_3$N(4--3) line emission distribution for the binary system 
    in Table\,\ref{tab:par} with the inclination of 0\arcdeg\ (face-on), 
    30\arcdeg, 50\arcdeg, 70\arcdeg, and 90\arcdeg\ (edge-on) from left 
    to right. The top panels display the images for the $-7.4\,\kmps$ 
    channel, while the middle and bottom panels show the P-V diagram along 
    $x$- and $y$-axis, respectively. The line of nodes of the binary 
    orbits ($x$-axis; major axis) corresponds to $\rm PA=10\arcdeg$ 
    in Figure\,\ref{fig:chm}. Attention is in the changes of position 
    angle of knotty structure in both channel image and P-V diagram 
    (marked by arrows), the tilt of pattern in the P-V diagram along 
    $y$-axis at mid-range inclinations, and the increase of maximum 
    velocity with inclination. The color ranges from 0.1 to 2\,mJy/beam 
    in a logarithmic scale from blue to red, for the synthetic beam size 
    of $\sim0.\!\!\arcsec7$ and velocity resolution of $\sim1\,\kmps$. 
  }
\end{figure*}

Figure\,\ref{fig:chm} shows the modeled HC$_3$N(4--3) line emission with 
the parameters in Table \ref{tab:par} in green color, overlaid with the 
observational data displayed in red color. A good correlation exhibits 
between the observed and modeled data in the individual velocity channels, 
albeit the outermost arm in the observation is barely detected at the 
$3\sigma$ level, where the noise level $\sigma=0.5$\,mJy/beam refers the 
root-mean-square of the line-free channels in the VLA observation. The 
observed channel images (red) range between 1 and 5\,mJy/beam. To present 
the pattern shape beyond the sensitivity limit of the observation, the 
modeled channel images (green) are displayed from 0.1 to 2\,mJy/beam.

All arc patches in the observed line emission are well located on 
the modeled spiral pattern. At the central channels (middle panels 
of Figure\,\ref{fig:chm}) the model reveals elongated spiral patterns, 
while at higher velocity channels (top and bottom panels) ring-like 
patterns are reproduced. Here, we emphasize that the ring-like features 
appear in the spiral-shell models with considerably large inclination 
angles \citep[see][]{kim12b}. It suggests that ring-like patterns 
observed in the AGB circumstellar envelopes and in the outer parts 
of PPN/PN do not necessarily imply an underlying spherically symmetric 
nature. The inclination angle of the presented model is 60\arcdeg. 

The binary model in Figure\,\ref{fig:chm} shows the breakup of rings 
and their unequal spacing between adjacent arms. This appears in the 
intermediate range of inclination angles, reflecting the transition from 
a spiral shape to a ring shape, as explained in Section \ref{sec:inc}. 
In the $-7.4\,\kmps$ channel, for example, the ring-like pattern is 
incomplete revealed by the gap in the east at $\rm PA=100\arcdeg$ 
(along the east side of minor axis; see also the first panel of 
Figure\,\ref{fig:pvd}). The ring radius of the northern part of the 
gap is larger than the radius of the southern part, in which the model 
and observation match sensibly. 

The companion's gravitational wake appears as knots in the modeled channel 
images located in specific PAs (marked by arrows in Figure\,\ref{fig:chm}) 
as determined by the inclination. In the zero-velocity channel (middle 
panel of Figure \ref{fig:chm}) representing the midplane, the knots are 
arranged along the line of nodes of the orbits (i.e., $\rm PA=10\arcdeg$ 
and 190\arcdeg). The knots shift to the east in the redshifted velocity 
channels and to the west in the blueshifted channels. The direction 
of shift of the knots with the channel velocity is determined by the 
orientation of the companion's orbit. The model knots eventually 
converge on the east and west ends of the ring pattern at the maximum 
velocity channels, which results in the thickening and brightening of 
the pattern at such locations.

The PAs of knots in the modeled channel images are consistent with 
bright emission spots in the corresponding VLA images. For example, 
the observed image (in Figure\,\ref{fig:chm}) exhibits thickening 
and brightening features in the ring pattern on the east and west 
ends at the $13.2\,\kmps$ and $-11.5\,\kmps$ channels, respectively. 
Another example is found at the southwest in the $-3.3\,\kmps$ and 
$-7.4\,\kmps$ channels, where the detected emission in the outermost 
ring (red) corresponds to the position of bright knotty structure in 
the model (green). At the locations of these knots, the arm (in red 
and green images) is found considerably inside the analytic Archimedes 
spiral model without considering the companion's gravitational wake 
(gray solid line). This decrease of arm radii is explained by 
\citet[Figure 9 and text]{kim12b}, which showed that a shock front 
(appearing as knots in the channel images) at the location of the 
companion's wake reduces the speed of the downstream flow and, 
therefore, the pattern propagation speed.

\section{DISCUSSION}\label{sec:dis}

We have shown that a numerical hydrodynamic radiative transfer model 
for CIT 6 demonstrates that the observed incomplete ring patches can 
be naturally understood within the framework of a single spiral-shell 
pattern induced by a binary motion. In contrast, the previous spherical 
shell model \citep{cla11} and incomplete shell model \citep{cha12} 
required the use of individual shell parameters without providing 
physical understanding for their variations. Moreover, the binary 
model reveals that the knotty structures result from the overlapping  
between a spiral-shell shaped pattern due to the orbital motion of the 
mass losing star and a more flattened spiral pattern onto the orbital 
plane due to the companion's motion. In this picture, the knots are 
aligned in a specific PA, which systematically changes with the velocity 
channel. This implies that such regularly-aligned knots are a natural 
consequence of the model.

The total flux integrated over the full simulation domain is related to the 
mass loss rate and the HC$_3$N abundance for a given temperature distribution. 
We have adopted the mass loss rate \oops{of $3.2\times10^{-6}\,\Mspy$} and 
\oops{relative} molecular abundance \oops{of $1.5\times10^{-6}$\ } from 
\citet{zha09}. Unlike \citet{zha09}, wh\oops{o assumed} the local thermal 
equilibrium with a single temperature, the temperature distribution in our 
model is determined by an adiabatic gas equation of state. The resulting 
total flux is smaller than the VLA observation by a factor of 2. 
\oops{This implies that a} complete modeling of these parameters requires 
studies involving a further fine-tuning of the input parameters as well as 
the introduction of molecular chemistry. Such studies will also necessitate 
incorporating observations of various molecular species, transitions, and 
angular resolutions. However, these parameters merely scale the flux level, 
but do not change the shape of the spiral pattern which is the focus of 
this paper.

The fact that the observed redshifted emission component is stronger than 
the blueshifted component is, however, not reproduced in our spiral model, 
albeit a small magnitude of spectral asymmetry is found depending on the 
inclination. This implies that our model is incomplete and that the asymmetry 
along the line-of-sight does not originate solely from the binary motion. 
The observed spectral asymmetry may agree with a presence of a bow shock due 
to the interaction between the circumstellar and interstellar media if the 
space motion away from us is significant \citep[e.g.,][]{wil96,war06,may11}. 
However, the VLA data does not show a shift of the Local Standard of Rest 
(LSR) velocity from the central velocity of the spiral-shell pattern. 
Other possible effects which may enhance the asymmetry include physical 
and chemical inhomogenities in the surrounding matter or in the dust 
formation phase at the onset of the stellar wind \citep[e.g.,][]{sok00,
woi05} if propagated to large scales. 

The missing emission in the west side of the central channels is also beyond 
the scope of this paper. This observed feature cannot be explained simply by 
a spiral-shell model nor by a spherical shell model. This spatial asymmetry 
is also found in a different transition of HC$_3$N and in its parent molecule 
CN \citep{lin00,din09}. We speculate that such lack of emission could possibly 
be caused by a chemical effect reducing the molecular abundance, rather than 
being affected by the gas density.

Compared to the other model sets in the parameter space determined 
by the projected pattern shape (Figure\,\ref{fig:deg}), for instance 
at $(V_1,\,\Vwav)=(3.65,\,15.7)$, the presented model shows a better 
match with the observation for the pattern features explained in 
Section \ref{sec:inc}--\ref{sec:res} in both channel maps and P-V 
diagrams. However, the difference of our fiducial model from the other 
comparison model sets are not decisive for the given uncertainties in 
the model parameters. In particular, observational uncertainties due to 
the limited sensitivity in determining the elongated shape and the arm 
spacing, as well as the determination of the source distance, stellar 
luminosity, photospheric temperature, and binary separation enter. 
Follow-up observations with a higher sensitivity for a chemically stable 
molecule such as CO would address some of these uncertainties.

\section{SUMMARY}\label{sec:sum}

We have developed a method to investigate the incomplete ring-like feature 
of a spiral-shell pattern in an outflowing circumstellar envelope, due to 
the orbital motion of central binary stars with an inclination \oops{with 
respect to the orbital plane}, providing an alternative explanation to models 
invoking periodic mass loss events of a single star. As a first step in this 
method, a parameter space study is performed as in \citet{kim12c} and improved 
upon by taking account of the inclination angle for the constraint on the wind 
expansion velocity. The results of such an analysis significantly narrow down 
the parameter space that can provide a fit to the observed pattern shape in 
the plane of the sky. This reduction in the parameter space is important for 
the second step where a modest number of hydrodynamic and radiative transfer 
investigations can be carried out. The pattern shape, degree of asymmetry, 
position of knots in individual velocity channels and P-V diagrams are used 
via detailed modeling to constrain binary properties. 

We have shown that gas kinematics is important in constraining the binary 
properties, which would otherwise remain degenerate under the constraints 
from the parameter space analysis based only on the shape of the circumstellar 
spiral-shell pattern in the plane of the sky. We have also demonstrated 
that a simple analytic analysis for the elongated spiral model based on 
an Archimedes spiral-shell pattern viewed as a function of inclination angle 
for comparison to individual velocity channels is useful for determining 
the applicability of the model to observational data. Finally, observed 
circumstellar patterns often reveal wiggles with varying slope in the polar 
coordinates that can be adequately described by the elongated spiral model, 
not by a face-on Archimedes spiral nor by spherical shells. 

We have used the AGB star, CIT 6, as the first example to illustrate 
\oops{a} method \oops{which can be used to help to constrain} the binary 
properties given the existence of high resolution and sensitivity VLA 
data for the HC$_3$N(4--3) molecular line. This first success of applying 
the binary-induced spiral-shell model to an observed incomplete ring-like 
pattern opens the possibility of connecting the ring-like patterns commonly 
found in the AGB circumstellar envelopes and in the outer parts of PPN/PN 
and pointing to the conceivable presence of central binary systems.

\acknowledgments

We acknowledge a detailed comments from an anonymous referee, and thank 
Dr.\ Mark Claussen for providing the calibrated FITS image of the VLA data. 
This research is supported by the Theoretical Institute for Advanced 
Research in Astrophysics (TIARA) in the Academia Sinica Institute of 
Astronomy and Astrophysics (ASIAA). The computations presented here 
have been performed through the ASIAA/TIARA computing resource. 
The SPARX code development is supported by the Coordinated/Coherent 
Hydrodynamic and Astrophysical Research, Modeling and Synthesis (CHARMS) 
project in ASIAA. 
The FLASH code is developed by the DOE-supported ASC/Alliance Center 
for Astrophysical Thermonuclear Flashes at the University of Chicago. 

\begin{deluxetable}{cccccccccccc}
\tablecolumns{12}
\tablecaption{Model parameters\label{tab:par}}

\tablehead{\colhead{$M_1$} & \colhead{$M_2$} & \colhead{$r_1+r_2$}
 & \colhead{$i$} & \colhead{$\Vwav$} & \colhead{}
 & \colhead{$T_{\rm in}$} & \colhead{$r_{\rm in}$} & \colhead{$\gamma$}
 & \colhead{} & \colhead{$\dot{M}_1$} & \colhead{HC$_3$N abundance\tablenotemark{\dag}}\\
\colhead{[\Msun]} & \colhead{[\Msun]} & \colhead{[AU]}
 & \colhead{[deg]} & \colhead{[\kmps]} & \colhead{}
 & \colhead{[$K$]} & \colhead{[AU]} & \colhead{}
 & \colhead{} & \colhead{[\Mspy]} & \colhead{[of H$_2$]}
}
\startdata
0.8 & 2.2 & 68 
& 60 & 15.7 & 
& 950 & 10 & 1.4 & 
& $3.2\!\times\!10^{-6}$ & $1.5\!\times\!10^{-6}$
\enddata
\tablecomments{Current mass of mass losing star $M_1$, mass of its 
companion $M_2$, their separation $r_1+r_2$, orbit inclination $i$, 
and intrinsic wind velocity from the mass losing star $\Vwav$ are 
chosen based on the parameter space analysis as indicated with the 
plus sign in Figure\,\ref{fig:deg}. The modeled binary stars are 
currently located at the line of node ($\phi_{\rm binary}=0\arcdeg$). 
$T_{\rm in}$ refers the temperature at reference position, $r_{\rm in}$, 
representing the wind formation radius. $\gamma$ is the adiabatic index 
of gas. The mass loss rate $\dot{M}_1$ and HC$_3$N abundance are taken 
from \citet{zha09}.}
\tablenotetext{\dag}{The abundance within 2\arcsec\ is set to be zero 
(See text).}
\end{deluxetable}

\begin{thebibliography}{}
\bibitem[Balick \& Frank(2002)]{bal02} Balick, B., \& Frank, A.\ 2002, \araa, 40, 439 
\bibitem[Binney \& Tremaine(2008)]{bin08} Binney, J., \& Tremaine, S.\ 2008, Galactic Dynamics (Princeton, NJ: Princeton Univ. Press)
\bibitem[Bond(1996)]{bon96} Bond, H.\ 1996, HST press release STScI-PR96-05 
\bibitem[Colella \& Woodward(1984)]{col84} Colella, P., \& Woodward, P.~R.\ 1984, Journal of Computational Physics, 54, 174
\bibitem[Chau et al.(2012)]{cha12} Chau, W., Zhang, Y., Nakashima, J.-i., Deguchi, S., \& Kwok, S.\ 2012, \apj, 760, 66 
\bibitem[Claussen et al.(2011)]{cla11} Claussen, M.~J., Sjouwerman, L.~O., Rupen, M.~P., et al.\ 2011, \apjl, 739, L5 
\bibitem[Cohen(1979)]{coh79} Cohen, M.\ 1979, \mnras, 186, 837 
\bibitem[Cohen \& Hitchon(1996)]{coh96} Cohen, M., \& Hitchon, K.\ 1996, \aj, 111, 962 
\bibitem[Decin et al.(2011)]{dec11} Decin, L., Royer, P., Cox, N.~L.~J., et al.\ 2011, \aap, 534, A1 
\bibitem[De Marco(2009)]{dem09} De Marco, O.\ 2009, \pasp, 121, 316 
\bibitem[Dinh-V.-Trung \& Lim(2009)]{din09} Dinh-V.-Trung, \& Lim, J.\ 2009, \apj, 701, 292 
\bibitem[Edgar et al.(2008)]{edg08} Edgar, R.~G., Nordhaus, J., Blackman, E.~G., \& Frank, A.\ 2008, \apjl, 675, L101
\bibitem[Fryxell et al.(2000)]{fry00} Fryxell, B,. Olson, K., Ricker, P., et al.\ 2000, \apjs, 131, 273
\bibitem[He(2007)]{he07} He, J.~H.\ 2007, \aap, 467, 1081
\bibitem[H{\"o}fner(2007)]{hof07} H{\"o}fner, S.\ 2007, Why Galaxies Care About AGB Stars: Their Importance as Actors and Probes, 378, 145 
\bibitem[Hogerheijde \& van der Tak(2000)]{hog00} Hogerheijde, M.~R., \& van der Tak, F.~F.~S.\ 2000, \aap, 362, 697 
\bibitem[Huggins(2007)]{hug07} Huggins, P.~J.\ 2007, \apj, 663, 342 
\bibitem[Huarte-Espinosa et al.(2013)]{hua13} Huarte-Espinosa, M., Carroll-Nellenback, J., Nordhaus, J., Frank, A., \& Blackman, E.~G.\ 2013, \mnras, 1448 
\bibitem[Hurley et al.(2000)]{hur00} Hurley, J.~R., Pols, O.~R., \& Tout, C.~A.\ 2000, \mnras, 315, 543 
\bibitem[Kim(2011)]{kim11} Kim, H.\ 2011, \apj, 739, 102 
\bibitem[Kim \& Taam(2012a)]{kim12a} Kim, H., \& Taam, R.~E.\ 2012a, \apj, 744, 136 
\bibitem[Kim \& Taam(2012b)]{kim12b} Kim, H., \& Taam, R.~E.\ 2012b, \apj, 759, 59 
\bibitem[Kim \& Taam(2012c)]{kim12c} Kim, H., \& Taam, R.~E.\ 2012c, \apj, 759, L22 
\bibitem[Le{\~a}o et al.(2006)]{lea06} Le{\~a}o, I.~C., de Laverny, P., M{\'e}karnia, D., de Medeiros, J.~R., \& Vandame, B.\ 2006, \aap, 455, 187 
\bibitem[Ladjal et al.(2010)]{lad10} Ladjal, D., Barlow, M.~J., Groenewegen, M.~A.~T., et al.\ 2010, \aap, 518, L141 
\bibitem[Le Bertre(1992)]{leb92} Le Bertre, T.\ 1992, \aaps, 94, 377 
\bibitem[Lindqvist et al.(2000)]{lin00} Lindqvist, M., Sch{\"o}ier, F.~L., Lucas, R., \& Olofsson, H.\ 2000, \aap, 361, 1036
\bibitem[Loup et al.(1993)]{lou93} Loup, C., Forveille, T., Omont, A., \& Paul, J.~F.\ 1993, \aaps, 99, 291 
\bibitem[Madej et al.(2004)]{mad04} Madej, J., Nale{\.z}yty, M., \& Althaus, L.~G.\ 2004, \aap, 419, L5
\bibitem[Maercker et al.(2012)]{mae12} Maercker, M., Mohamed, S., Vlemmings, W.~H.~T., et al.\ 2012, \nat, 490, 232 
\bibitem[Mastrodemos \& Morris(1999)]{mas99} Mastrodemos, N., \& Morris, M.\ 1999, \apj, 523, 357 
\bibitem[Mattsson et al.(2007)]{mat07} Mattsson, L., H{\"o}fner, S., \& Herwig, F.\ 2007, \aap, 470, 339 
\bibitem[Mauron \& Huggins(2006)]{mau06} Mauron, N., \& Huggins, P.~J.\ 2006, \aap, 452, 257 
\bibitem[Mayer et al.(2011)]{may11} Mayer, A., Jorissen, A., Kerschbaum, F., et al.\ 2011, \aap, 531, L4 
\bibitem[Monnier et al.(2000)]{mon00} Monnier, J.~D., Tuthill, P.~G., \& Danchi, W.~C.\ 2000, \apj, 545, 957 
\bibitem[Morris(1987)]{mor87} Morris, M.\ 1987, \pasp, 99, 1115 
\bibitem[Nordhaus \& Blackman(2006)]{nor06} Nordhaus, J., \& Blackman, E.~G.\ 2006, \mnras, 370, 2004 
\bibitem[Paczy{\'n}ski(1971)]{pac71} Paczy{\'n}ski, B.\ 1971, \actaa, 21, 271 
\bibitem[Sahai et al.(2011)]{sah11} Sahai, R., Morris, M.~R., \& Villar, G.~G.\ 2011, \aj, 141, 134 
\bibitem[Sahai et al.(1998)]{sah98} Sahai, R., Trauger, J.~T., Watson, A.~M., et al.\ 1998, \apj, 493, 301 
\bibitem[Schmidt et al.(2002)]{sch02} Schmidt, G.~D., Hines, D.~C., \& Swift, S.\ 2002, \apj, 576, 429 
\bibitem[Sch{\"o}ier et al.(2005)]{sch05} Sch{\"o}ier, F.~L., van der Tak, F.~F.~S., van Dishoeck, E.~F., \& Black, J.~H.\ 2005, \aap, 432, 369 
\bibitem[Soker(1994)]{sok94} Soker, N.\ 1994, \mnras, 270, 774
\bibitem[Soker(2000)]{sok00} Soker, N.\ 2000, \mnras, 312, 217 
\bibitem[Su(2004)]{su04} Su, K.~Y.~L.\ 2004, Asymmetrical Planetary Nebulae III: Winds, Structure and the Thunderbird, 313, 247 
\bibitem[Theuns \& Jorissen(1993)]{the93} Theuns, T., \& Jorissen, A.\ 1993, \mnras, 265, 946 
\bibitem[Trammell et al.(1994)]{tra94} Trammell, S.~R., Dinerstein, H.~L., \& Goodrich, R.~W.\ 1994, \aj, 108, 984 
\bibitem[Trammell \& Goodrich(1997)]{tra97} Trammell, S.~R., \& Goodrich, R.~W.\ 1997, AAS, 191, 47.15 
\bibitem[Wareing et al.(2006)]{war06} Wareing, C.~J., Zijlstra, A.~A., Speck, A.~K., et al.\ 2006, \mnras, 372, L63 
\bibitem[Wilkin(1996)]{wil96} Wilkin, F.~P.\ 1996, \apjl, 459, L31 
\bibitem[Willson \& Hill(1979)]{wil79} Willson, L.~A., \& Hill, S.~J.\ 1979, \apj, 228, 854 
\bibitem[Woitke \& Niccolini(2005)]{woi05} Woitke, P., \& Niccolini, G.\ 2005, \aap, 433, 1101 
\bibitem[Wood(1979)]{woo79} Wood, P.~R.\ 1979, \apj, 227, 220 
\bibitem[Zhang et al.(2009)]{zha09} Zhang, Y., Kwok, S., \& Dinh-V-Trung 2009, \apj, 691, 1660 
\end{thebibliography}
\end{document}